\documentclass[aps,twocolumn,superscriptaddress,showpacs]{revtex4}
 \usepackage{amsmath}
 \usepackage{amssymb}
 \usepackage{amsfonts}
 \usepackage{graphicx}
 \usepackage{multirow}
 \usepackage{bm}
 \usepackage{subfigure}
 \usepackage{enumerate}
 \usepackage{dcolumn}
 \usepackage{bm}
 \usepackage{graphicx}
 \usepackage{color}
 \DeclareGraphicsExtensions{. jpg,. pdf, . mps, . png, . eps, . ps, . EPS}

\begin{document}
 \def \be{ \begin{equation}}
 \def \ee{ \end{equation}}
 \def \bc{ \begin{center}}
 \def \ec{ \end{center}}
 \def \bea{ \begin{eqnarray}}
 \def \eea{ \end{eqnarray}}
 \newcommand{ \avg}[1]{ \langle{#1} \rangle}
 \newcommand{ \Avg}[1]{ \left \langle{#1} \right \rangle}

 \title{Monochromaticity in Neutral Evolutionary Network Models}

 \author{Arda Halu}
 \affiliation{Department of Physics, Northeastern University, Boston 02115 MA, USA}
 \author{Ginestra Bianconi}
 \affiliation{Department of Physics, Northeastern University, Boston 02115 MA, USA}
 \begin{abstract}
Recent studies on epistatic networks of model organisms have unveiled a certain type of modular property called monochromaticity in which the networks are clusterable into functional modules that interact with each other through the same type of epistasis. Here we propose and study three epistatic network models that are inspired by the Duplication-Divergence mechanism to gain insight into the evolutionary basis of monochromaticity and to test if it can be explained as the outcome of a neutral evolutionary hypothesis. We show that the epistatic networks formed by these stochastic evolutionary models have monochromaticity conflict distributions that are centered close to zero and are statistically significantly different from their randomized counterparts. In particular, the last model we propose yields a strictly monochromatic solution. Our results agree with the monochromaticity findings in real organisms and point toward the possible role of a neutral mechanism in the evolution of this phenomenon.    
 \end{abstract}
 \maketitle
\section{Introduction}
Highly interacting molecular networks \cite{Barabasi, Toroczkai, Sneppen_book, Bornholdt, Reka, Kepes, Alon} are at the forefront of systems biology and are essential to understand biological systems beyond the single molecule framework. In these networks usually a modular  topology \cite{Alon, Modular} is predictive of the presence of  functional modules and therefore topological considerations \cite{Santo, PNAS} have  relevance in detecting the function and dynamics of these networks. The origin of the functional modularity is not thoroughly characterized: at first it was considered to be due to evolutionary pressure \cite{Alon} but more recently observations have shown that modularity might be the natural outcome of neutral evolution \cite{Sole_modular}.
 
Recently the attention has been directed to the study of complex genetic interactions \cite{segre, costanzo}.
In fact the  relationship between phenotypes and genotypes is a complex one -- phenotypes are most commonly determined by simultaneous interactions between multiple genes rather than by a single gene \cite{phillips}. These genetic interactions are furthermore complicated by the fact that the effects of individual genes are modified by other genes, resulting in the phenomenon called epistasis. 
The epistatic network between mutations or genetic variations in the genome  is found to
play a crucial role in determining the fitness of different organisms \cite{segre,costanzo}. This network is interacting with the other biological networks of the cell such as the protein-protein interaction network\cite{Uetz,Sneppen} the metabolic network\cite{Laszlo_met} or the transcription network \cite{Kepes, Marco} and is crucial in determining the phenotype of an organism. The degree distribution of these networks is clearly fat-tailed \cite{costanzo} thus showing a similar topology with respect to other biological networks. Moreover the genes involved in epistatic interactions can have a hierarchical relationship; they can mask or modify each other's effects and even combine to create new phenotypes.

 A mathematical definition of epistasis that was introduced by population geneticists in order to deal with complex traits such as human diseases in a quantitative and statistical way is known as the multiplicative model. Widely used in genetic interaction studies, it makes use of the quantitative measure $\epsilon$ which is the difference between the fitness $W_{AB}$ of a double mutant and the product of fitnesses of two single mutants $W_A, W_B$, i.e.
\begin{equation*}
\epsilon = W_{AB} - W_A W_B
\end{equation*}
\noindent The sign of this deviation from multiplicative behavior helps distinguish between two different classes of interactions: a positive (antagonistic) epistasis signifies a buffering type of interaction in which the effect of the double mutation is less severe than single mutations combined; a negative (synergistic) epistasis means an aggravating type of interaction where the double mutation results in a more severe effect on the fitness. 

Biological systems are known to often display a functionally modular architecture on various levels \cite{hartwell}. Much of the robustness and adaptability of biological networks has been attributed to this modular structure. Modularity in biological networks has been studied from the perspectives of topology \cite{ravasz}, information theory \cite{ziv} and more recently, epistasis. 

A recent study \cite{segre} has investigated epistasis through a genetic interaction network constructed by the calculation of epistasis values $\epsilon$ from the fitnesses of all of the single and double knockouts of 890 metabolic genes of {\it Saccharomyces Cerevisiae} (budding yeast). In both supervised organization with known gene annotations and unsupervised organization with unknown gene functions, the study revealed a structural clusterability into modules that interact with each other monochromatically, that is, via the same type of epistasis. Monochromaticity in the interactions between functional modules was later verified as part of a genome-wide genetic interaction study \cite{costanzo} of {\it Saccharomyces Cerevisiae}. In this larger scale research, the genetic interaction map of the budding yeast was constructed from 5.4 million gene pairs. Modules belonging to different pathways and complexes were shown to be interacting via the same type of epistasis almost exclusively. Most recently, there have been efforts in introducing measures to quantify monochromaticity \cite{baryshnikova, hsu}. 


In this work, we are motivated by these findings to ask how the monochromaticity feature emerged in nature. We want to gain insight into whether it can be explained as the outcome of a neutral evolutionary theory, driven by stochastic processes. To this end we devise network models inspired by the duplication-divergence model \cite{sole, vazquez, kim, wagner} that is shown to successfully capture the heterogeneous network properties and other important topological features of protein interaction networks. Focusing on the genomic level, we propose neutral evolutionary models in which epistatic networks of genes are constructed via similar duplication-divergence processes for the epistatic interaction signs. Signs of links are copied as genes are duplicated and they are switched and/or rewired with finite probability. We generate ensembles of epistatic networks using these models and assess their monochromaticity using the PRISM algorithm \cite{segre} which makes use of agglomerative clustering to detect modules. The number of monochromaticity violations $Q_{module}$ of these networks are compared to their randomized versions where the topology is preserved and the epistatic signs are shuffled. We show that the $Q_{module}$ distribution for epistatic network ensembles generated by these models are centered around zero indicating monochromaticity, and are statistically significantly different from their randomized counterparts, which are not monochromatic. In particular, the last model that we propose in which we assume  epistasis between genes that encode proteins belonging to protein complexes,  displays a clear separation of its conflict number distribution from that of the randomized ones and is strictly monochromatic. The degree distributions of each of these models display a fat tailed behavior, with the degree distributions of positive and negative interactions showing no deviation from each other as there is no bias towards any sign in any of the models.


This paper is structured as follows: In Section II we briefly revise the literature on duplication-divergence (D-D) models, summarize the PRISM algorithm and describe in detail the D-D inspired evolutionary models that we use to construct the epistatic networks. We study the monochromaticity conflict distributions of ensembles of networks produced by these models and their randomized counterparts. We finally make our concluding remarks in Section III.

\section{Monochromaticity in Duplication-Divergence Inspired Epistatic Models}
Gene duplication and the subsequent mutations (divergence) is believed to be one of the most crucial mechanisms driving evolution \cite{ohno}. In the past decade, mathematical models that make use of the duplication-divergence mechanism have been proposed and studied analytically \cite{sole, vazquez, kim, wagner}. These models were successful in capturing the basic topological properties of protein intraction networks such as scale-free degree distribution, small-world properties, modular architecture and robustness against random node removal. The duplication parts of these models are carried out in a similar way to each other. First, a random gene is selected to be duplicated, which is called a target gene. The duplicated new gene then acquires all of the interactions of the target gene. This makes sense in the proteomic viewpoint as the interactions of proteins are determined by their structure, which remains unchanged after duplication. The divergence part, on the other hand, is model specific and shows some variation from model to model in terms of the exact divergence mechanism. Usually, what is understood by divergence is the removal of interactions with some finite probability. Duplication-divergence models are divided into two classes according to the link removal mechanism. In asymmetric models \cite{sole, krapivsky} only the duplicated genes can lose links whereas in symmetric duplication-divergence models \cite{vazquez}, both the original and duplicated genes can lose their interactions. In some of the duplication-divergence models, this removal can be accompanied by rewiring of the links or formation of new links with some other finite probability.

In this paper we implement three epistatic network models based on the duplication-divergence mechanism with additional focus on the interaction signs. The epistatic interactions are mediated by protein interaction networks. Therefore a duplication-divergence model for the evolution of epistatic interactions implements a minimal hypothesis on their dynamical evolution. We start from a simple initial condition common to all of the three models. During the duplication phase, the interaction type is preserved whereas in the case of rewiring, the epistatic signs of the new links are chosen as detailed in the following subsections for each model. 

The monochromaticity -- the feature of clusterability into modules that interact via the same type of epistasis -- of each model is studied using the PRISM algorithm \cite{segre}, which was originally developed to study the yeast metabolic network. Belonging to a class of hierarchical clustering methods known as agglomerative clustering, the algorithm starts out with each element in a cluster of its own. It assigns a conflict score $C_{x, y}$ to every cluster pair $(x, y)$, which is total number of ``mixed sign'' or nonmonochromatic links and then at each step determines the set of cluster pairs with the minimum conflict score
\begin{equation*}
C_m = \min_{x, y} \{C_{x, y}\}
\end{equation*} 
It then selects from this set the one pair of clusters with the highest proximity (determined by the density of links between them) and combines them. This process is repeated until the whole network is covered. Finally, a total monochromaticity violation number $Q_{module}$ which is the sum of the minimum conflict scores over all the steps
\begin{equation*}
Q_{module} = \sum{C_m}
\end{equation*}
is assigned to the final clustering solution. A total conflict number of zero means a fully monochromatic solution. The result of this analysis on the yeast metabolic network reported in \cite{segre} is that the unsupervised organization of the real yeast metabolic network yielded a fully monochromatic solution. Unsupervised in this context means having no prior knowledge of the genes' functional annotations. Moreover, this unsupervised organization of the epistatic network resulted in modules that agree well with existing gene annotations. A supervised version of this analysis was also carried out in this study for pre-assigned functional annotation groups, which also yielded a statistically significant enrichment of monochromatic interactions.

In this paper we take a similar approach in which we investigate the distributions of the total conflict score $Q_{module}$ for the networks constructed by duplication-divergence models and their randomized counterparts. The color randomization scheme consists of the pairwise flipping of signs so that the topology of the network is preserved as well as the total number of positive and negative interactions. Following the unsupervised approach, we don't make any assumptions as to the function of genes. Since the duplication-divergence models are themselves of stochastic nature, we represent their total conflict score as a distribution as well.

\subsection{Model A}
\begin{figure}[h]
\begin{centering}
\includegraphics[width=0.5\columnwidth]{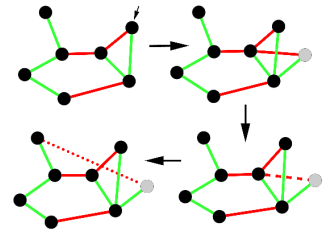}
\caption{(Color online) Duplication-Divergence scheme of the Model A. The target node is indicated by the small arrow and the duplicate node is in grey. Green (light grey) links denote positive interactions and red (dark grey) links negative interactions. Interactions are duplicated with the same signs; removed (dashed lines) with probability $\delta$ and linked (dotted lines) to previous nodes with random signs with probability $\alpha$.}
\label{sole}
\end{centering}
\end{figure}

The first model that we consider for the neutral evolution of epistatic networks is a simple generalization of the model introduced by Sol\'{e} et al. \cite{sole}. The Sol\'e model is one of the first models to be developed as a growing network model to simulate proteome evolution. It has two free parameters $\delta$ and $\alpha$ which are the probability of removal of the links of the duplicated gene and the probability of the duplicated gene to form links with any of the previously unlinked genes, respectively. It was shown to successfully capture structural properties of the yeast proteome such as the degree distribution \cite{sole}, sparse nature and modularity \cite{sole2}. 
Here we generalize this model in order to capture a simple neutral evolution of the epistatic network.
We start with simple initial conditions: at the beginning  the network is initialized with three nodes and two links so that $E_{12} = 1$ and $E_{13} = -1$ where E is the interaction matrix composed of the elements $-1$ (negative epistasis), 1 (positive epistasis) and 0 (no interaction). 
The epistatic implementation of this model for a network size of {\em n} consists of the iteration of the following set of rules (Fig. \ref{sole}). \\

\begin{enumerate}[(i)]

\item {\em duplication:} A target node $v_i \in V$ is chosen at random and the replicated node $v_r = v_{n+1}$ acquires all the links $\{e_{i, j}\}$ of the target node where $\{v_j\}$ is the set of neighbors of $v_i$. The signs are copied during duplication so that $E_{rj} = E_{ij}$.

\item {\em divergence 1 (deletion):} Each of the links $e_{r, j}$ of the replicated node is deleted with probability $\delta$.

\item {\em divergence 2 (addition of new links):} The replicated node $v_r$ is connected with a random sign to all of the previously unconnected nodes with probability $\alpha$, which is typically a small number.
\end{enumerate}

\noindent The parameters used in this work, namely $\delta = 0.5$ and $\alpha = 0.0002$ were chosen so that they are compatible with the estimated average connectivity from mean-field calculations \cite{sole} and experimental data.

\begin{figure}[h]
\begin{centering}
\vspace{-3mm}
\includegraphics[width=0.5\columnwidth]{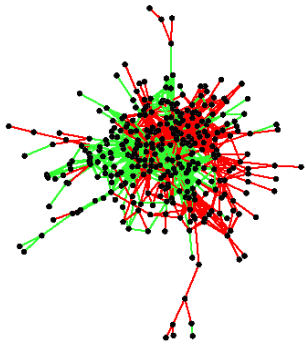}
\caption{(Color online) A sample network given by the  model A for $\delta = 0.5$ and $\alpha = 0.0002$. Positive and negative links are denoted by green (light grey) and red (dark grey) links, respectively.}
\label{solepic}
\end{centering}
\end{figure}

\begin{figure}[h]
\centering
\mbox{\hspace{-5mm} 
\subfigure{\includegraphics[width=0.27\textwidth]{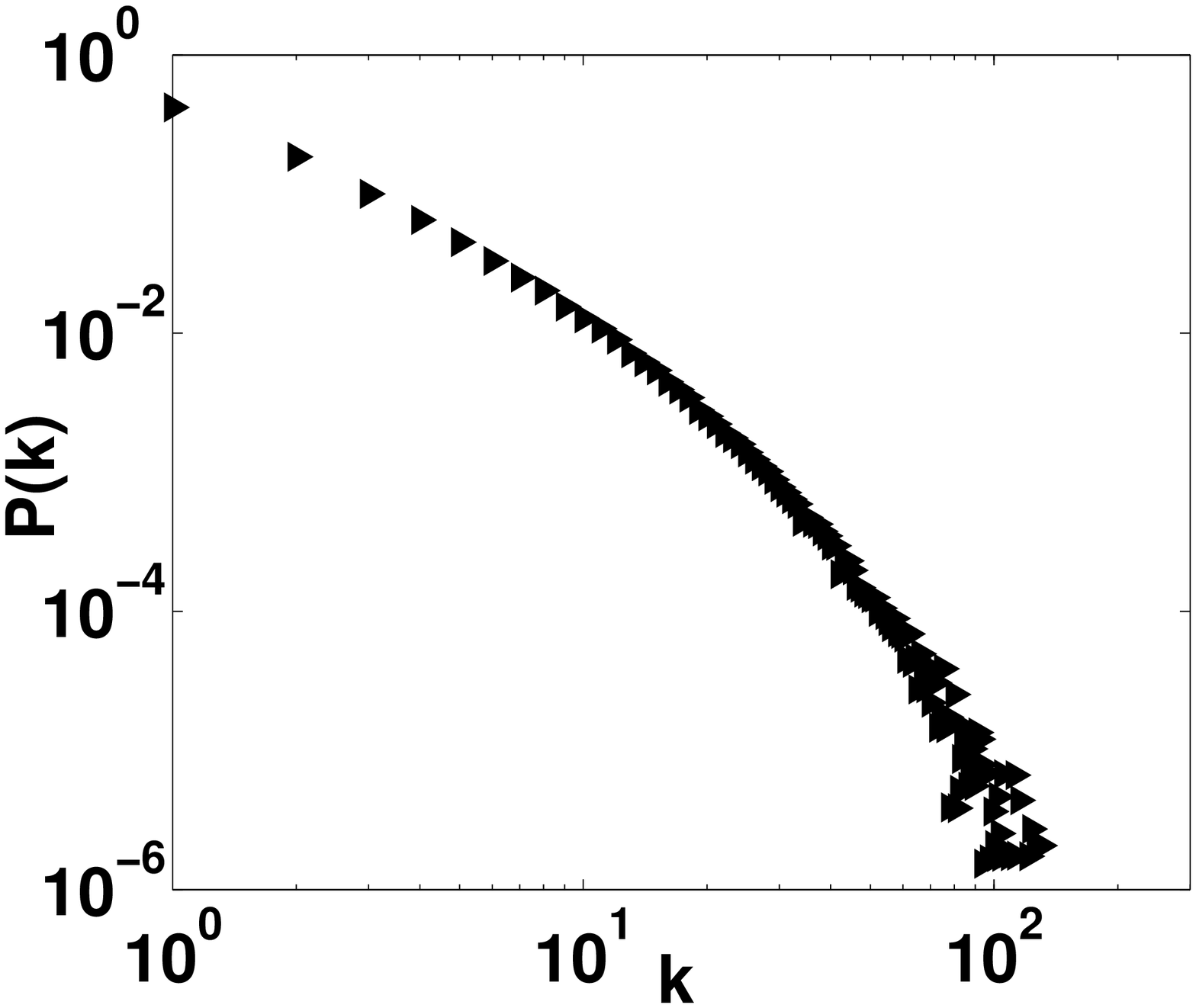}}
\hspace{-5mm}
\subfigure{\includegraphics[width=0.27\textwidth]{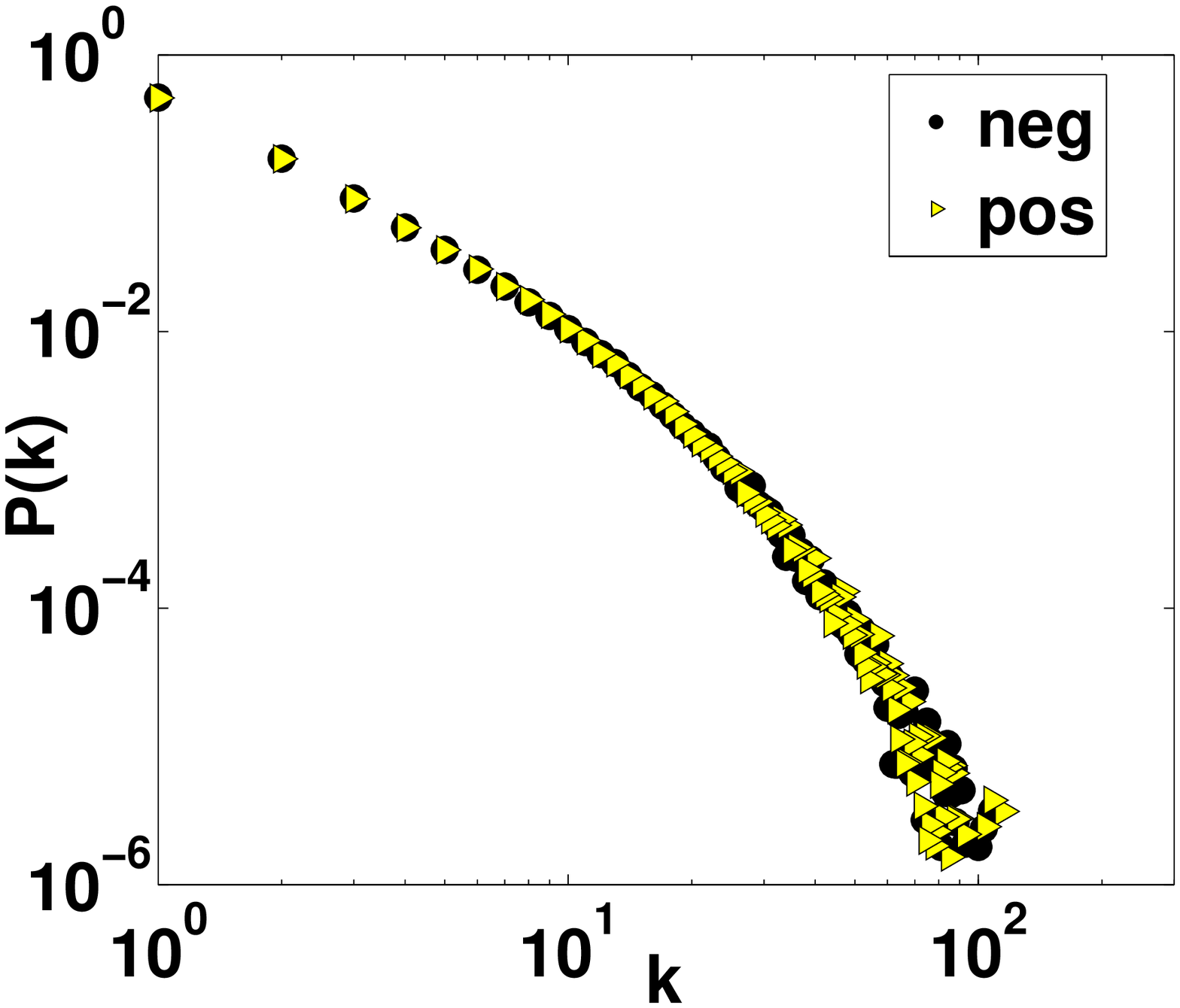}}}
\caption{(Color online) The total (left) and sign (right) degree distributions of the model A for $\delta = 0.5$ and $\alpha = 0.0002$. Circles and triangles denote negative and positive links, respectively.}
\label{soledist}
\end{figure}

An instance of the epistatic network generated by Model A is shown in Fig. \ref{solepic}. The total degree distribution and sign degree distributions (Fig. \ref{soledist}) of this model both display fat tailed distributions with an exponential cutoff although it is easy to disrupt the power law distribution of this model by increasing the rewiring probability, in which case the distribution becomes exponential similar to that of a random growing network. Furthermore, both positive and negative interaction degree distributions follow the same trend as there is no bias in the model towards any sign. 

\begin{figure}[h]
\begin{centering}
\vspace{-3mm}
\includegraphics[width=0.6\columnwidth]{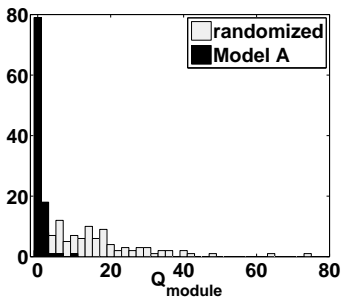}
\caption{The total monochromaticity violation number $Q_{module}$ distributions of model A networks (black) and their randomized counterparts (white) for 100 realizations of each. The parameters used are $\delta = 0.5$ and $\alpha = 0.0002$.}
\label{soleconflict}
\end{centering}
\end{figure}

The distribution of the total number of conflicts of the networks generated by Model A and their random counterparts is shown in Fig. \ref{soleconflict}. To get this distribution, a total number of 100 networks were generated and subsequently randomized. The resulting histogram shows a clear separation of the means in the duplication-divergence ensemble (shown in black) and randomized ensemble (shown in white). The conflict score of model A is very close to zero ($0.91 \pm 1.45$) whereas the corresponding randomized distribution has much larger spread. To test for the null hypothesis that the means of the distributions are the same, or in other words the null assumption that the expected $Q_{module}$ is given by the randomized network, we ran a statistical significance test on the distributions, which yielded a p-value of $8.43 \times 10^{-25}$, which means that the distributions are statistically significantly different, with model A giving a monochromatic solution whereas the randomized version isn't.

\subsection{Model B}

\begin{figure}[h]
\begin{centering}
\includegraphics[width=0.5\columnwidth]{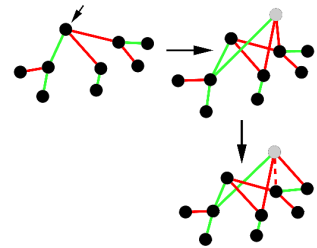}
\caption{(Color online) Duplication-Divergence scheme of model B. The target node is indicated by the small arrow and the duplicate node is in grey. Green (light grey) links denote positive interactions and red (dark grey) links negative interactions. Interactions are duplicated with the same signs; removed (dashed lines) with probability p and in the case of removal, rewired to one of the second neighbors of the target node with probability q according to the sign convention described below.}
\label{modelA}
\end{centering}
\end{figure}

In Model A the epistatic interactions of different signs remain strongly localized in different regions of the network as it appears clearly in the graphical representation of the network (Fig. \ref{solepic}).
In this section we provide a first modification of the model which has the effect of maintaining the fat-tail degree distribution while at the same time mixing further the epistatic interactions of different signs.
Model B  has two free parameters and takes into account the probability to connect to one of the second neighbors of the target node in the case of removal of the duplicated node. The duplication (i) step is the same as in the previous model whereas the divergence is different. 
Here in the following we give the precise definition of the algorithm we have implemented for Model B. The iterative steps are summarized in Figure \ref{modelA}.
We start with the same initial conditions as in model A, i.e. we start with three nodes and two links such that the epistatic interactions are given by $E_{12}=1$ and $E_{13}=-1$ where $E$ indicates the sign of the epistatic interaction.

\begin{enumerate}[(i)]

\item {\em duplication:} A target node $v_i \in V$ is chosen at random and the replicated node $v_r = v_{n+1}$ acquires all the links $\{e_{i, j}\}$ of the target node where $\{v_j\}$ is the set of neighbors of $v_i$. The signs are copied during duplication so that $E_{rj} = E_{ij}$.

\item {\em divergence 1 (deletion):} Each of the links $e_{r, j}$ of the replicated node is deleted with probability $p$.

\item {\em divergence 2 (rewiring):} The replicated node $v_r$ is connected to one of the second neighbors of the target node with probability $q$ in the case of removal of the duplicated link. The sign of the rewired link to the second neighbor is the product of the signs of the first and second neighbors such that $E_{rk} = E_{ij} \times E_{jk}$.
\end{enumerate}

\begin{figure}[h!]
\begin{centering}
\vspace{-3mm}
\includegraphics[width=0.5\columnwidth]{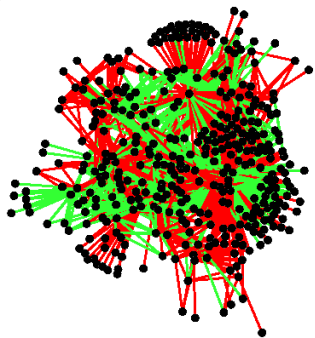}
\caption{(Color online) A sample network given by model B for $p = 0.4$ and $q = 0.01$. Positive and negative links are denoted by green (light grey) and red (dark grey) links, respectively.}
\label{modelApic}
\end{centering}
\end{figure}

\begin{figure}[h]
\centering
\mbox{\hspace{-5mm}
\subfigure{\includegraphics[width=0.27\textwidth]{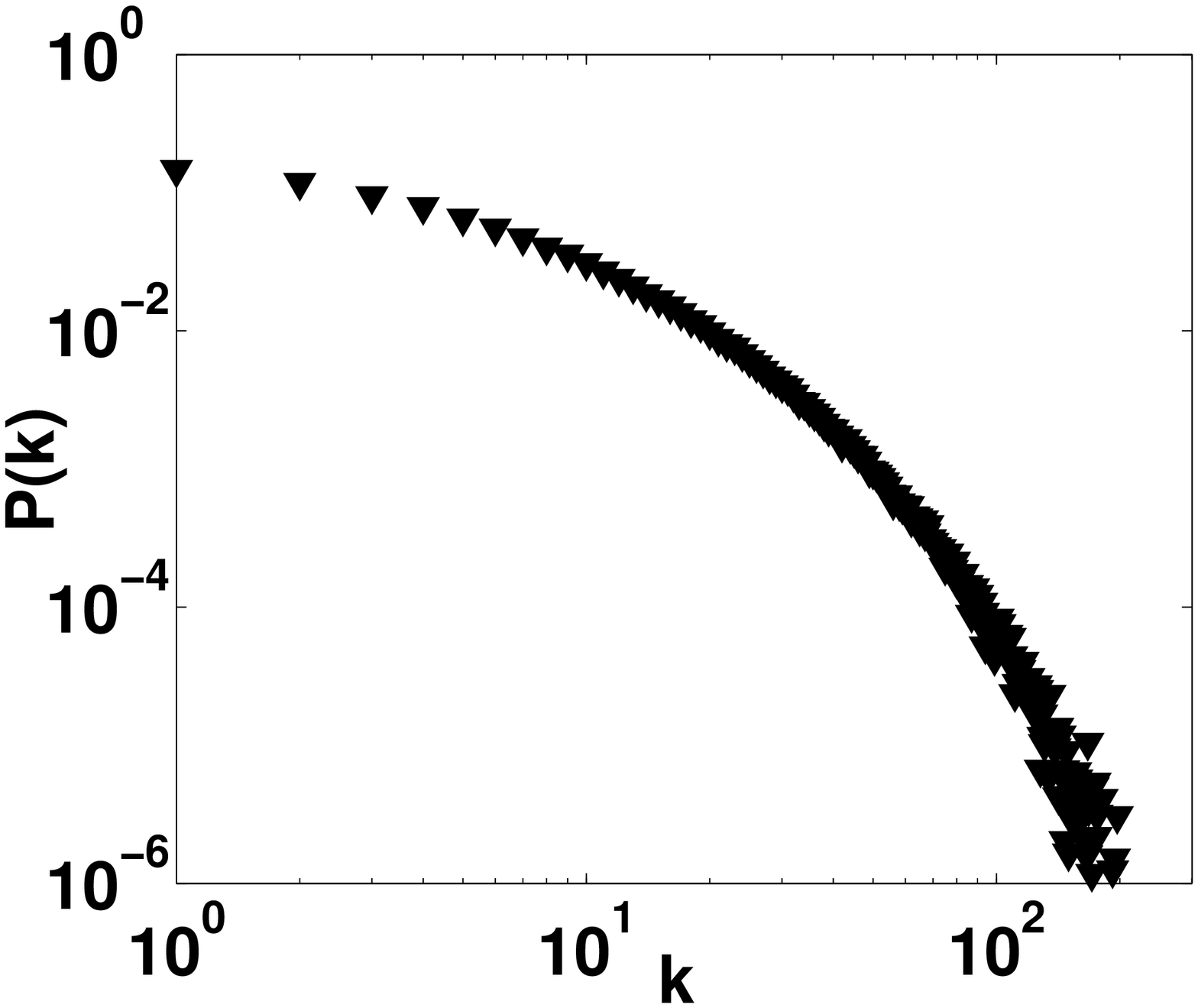}}
\hspace{-5mm}
\subfigure{\includegraphics[width=0.27\textwidth]{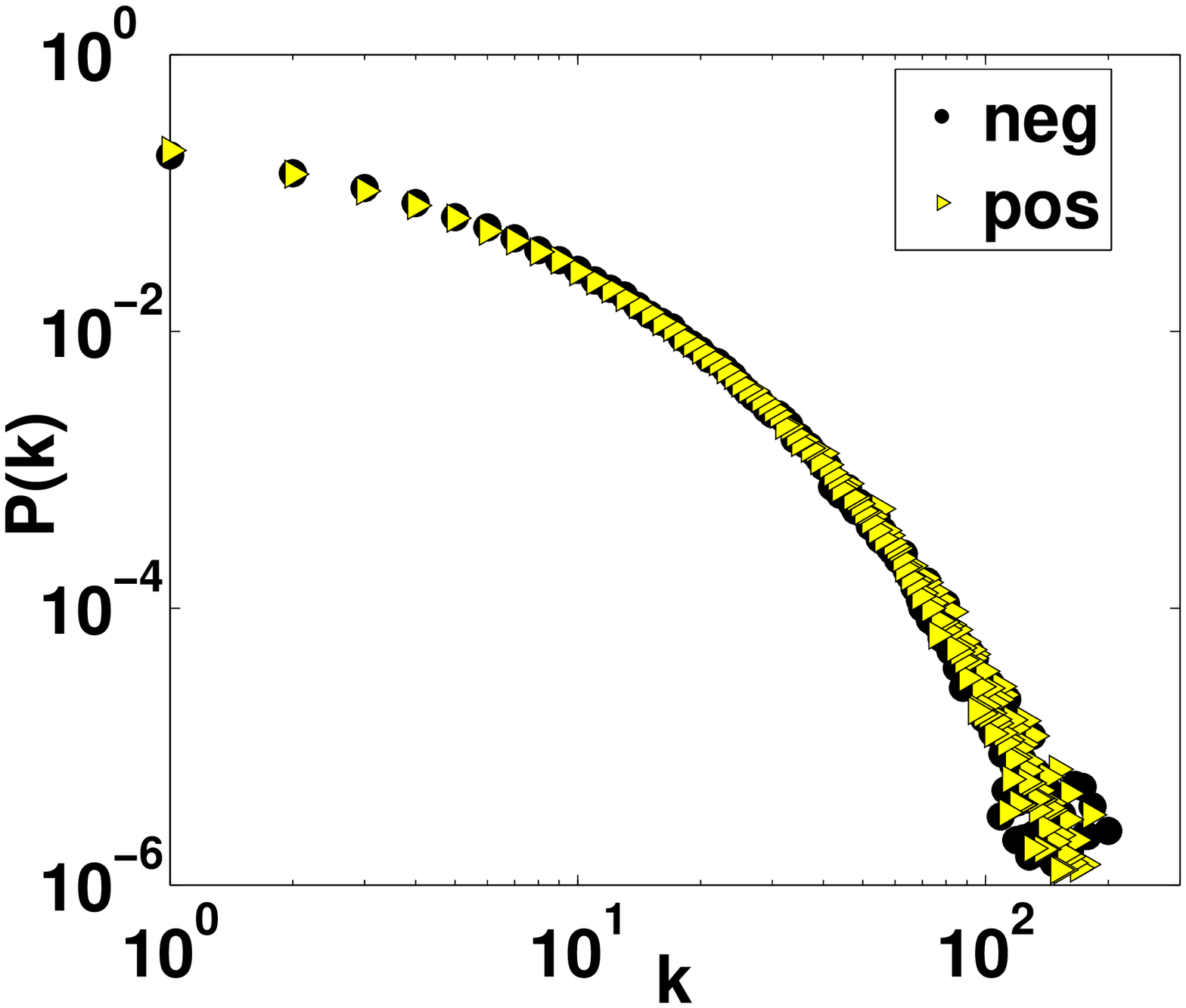}}}
\caption{(Color online) The total (left) and sign (right) degree distributions of model B for $p = 0.4$ and $q = 0.01$. Circles and triangles denote negative and positive links, respectively.}
\label{modelAdist}
\end{figure}

This model suggests that when a gene is duplicated, newly emerging epistatic interactions follow a transitive sign rule (divergence 2 mechanism). Although epistatic interactions are known to be non-linear and in general a transitivity of the sign might not be the rule, it is possible that for epistatic interactions mediated by protein-interaction networks this mechanism for divergence might effectively take place. 
A sample network generated by Model B is given in (Fig. \ref{modelApic}). The free parameters for this model were chosen as $p = 0.4$ and $q = 0.01$ for removal and rewiring, respectively. Here the network has a more mixed topology due to the rewiring to the second neighbors, therefore the monochromatically clusterable architecture is not evident from the figure of the network. Still, the scale free degree distribution is preserved (Fig. \ref{modelAdist}) as the rewiring is made with only one of the second neighbors of the target node, per removal of the duplicate link.  Positive and negative degree distributions are again indiscernible.

\begin{figure}[h]
\begin{centering}
\vspace{-3mm}
\includegraphics[width=0.6\columnwidth]{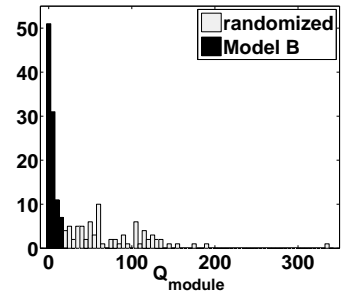}
\caption{The total monochromaticity violation number $Q_{module}$ distributions of model B networks (black) and their randomized counterparts (white) for 100 realizations of each. The parameters used are $p = 0.4$ and $q = 0.01$}
\label{Aconflict}
\end{centering}
\end{figure}

Duplication-divergence model B and randomized networks have the $Q_{module}$ distribution as shown in (Fig. \ref{Aconflict}). An ensemble  of 100 networks was used to collect statistics on monochromaticity for this model. In Figure $\ref{Aconflict}$ the total conflict distribution of model B is presented and it appears as slightly more spread than Model A although it is still centered close to zero ($3.83 \pm 4.36$). The corresponding randomized distribution is also more separated ($61.41 \pm 53.22$). The p-value under the same null hypothesis as the Model A is $1.26 \times 10^{-21}$. This also complies with the results above, indicating monochromatic separability of model B.

\subsection{Model C}

\begin{figure}[h]
\begin{centering}
\includegraphics[width=0.8\columnwidth]{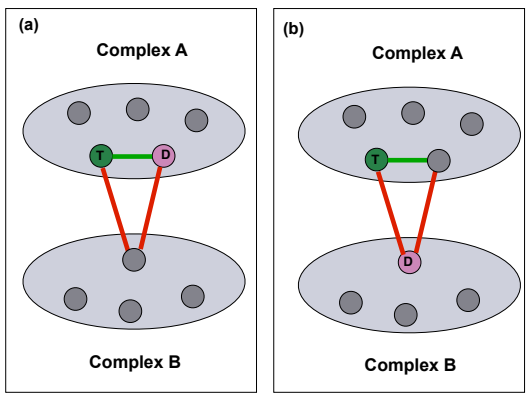}
\caption{(Color online) Schematic view of protein complexes and their epistatic interactions. A duplicated gene (D) can encode for a protein in the same protein complex as the template  gene (T) (Panel (a)),  or belong to a competing protein complex (Panel (b)). The figure shows the epistatic interactions in both cases.}
\label{pc}
\end{centering}
\end{figure}

This model is a generalization of the basic model which preserves the fat-tail degree distribution and implies a well-mixed  distribution of epistatic interactions with different signs.
The model is a very stylized one that takes into account the functional dependencies of protein complexes on each other. In this model, we assume that a duplicated gene might encode a protein that belongs to a protein complex which is either the same complex as that of the template protein (gene) or a complex that competes with the original template gene by aggravating interactions. Genes encoding proteins in competing protein complexes have aggravating (negative) interactions between them, pointing to a redundancy in the essential cellular functions of the two complexes. On the other hand, genes encoding proteins in the same complex have alleviating interactions between them. \cite{bandyopadhyay,casey}.   In Figure $\ref{pc}$ we give a schematic view of the two different possibilities that are contemplated in this model.

\begin{figure}[h]
\begin{centering}
\includegraphics[width=0.5\columnwidth]{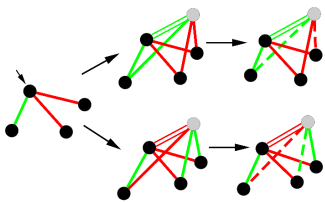}
\caption{(Color online) Duplication-Divergence scheme of model C. The template node is indicated by the small arrow and  the duplicate node is in grey. With equal probability  (1/2) the duplicated gene encodes for a protein in the same complex (double green (light grey) link) or in a competing complex (double red (dark grey) link)  with respect to the protein encoded by the target gene. Signs of the duplicated links are chosen according to the definition of the model. Removed interactions (with probability p) are shown by dashed lines.}
\label{modelB}
\end{centering}
\end{figure}
We start with the same initial conditions as in model A and B, i.e. we start with three nodes and two links such that the epistatic interactions are given by $E_{12}=1$ and $E_{13}=-1$ where $E$ indicates the sign of the epistatic interaction. At each iteration we perform the subsequent steps  (Fig. \ref{modelB}):

\begin{enumerate}[(i)]

\item {\em duplication:} A new duplicated node $v_r = v_{n+1}$ is a gene that encodes for a protein in the same complex ($E_{ri} = 1$) or in a competing complex  ($E_{ri} = -1$) with respect to that of a randomly chosen template node $v_i \in V$ with equal probability. If the replicated gene encodes for a protein in the same complex, the replicated node acquires all the links $\{e_{i, j}\}$ of the target node with the same sign, i.e. $E_{rj} = E_{ij}$. If the replicated gene encodes for a protein that belongs to a competing complex, the replicated node acquires all the links $\{e_{i, j}\}$ of the target node with the opposite sign, i.e. $E_{rj} = -E_{ij}$.  

\item {\em divergence (deletion):} Each of the links $e_{r, j}$ of the replicated node is deleted with probability $p$.

\end{enumerate}

\begin{figure}[h]
\begin{centering}
\includegraphics[width=0.2\columnwidth]{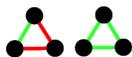}
\caption{(Color online) The two triangular motifs generated by model C. Positive and negative links are denoted by green (light grey) and red (dark grey) links, respectively.}
\label{Bmotifs}
\end{centering}
\end{figure}

\begin{figure}[h]
\begin{centering}
\vspace{-3mm}
\includegraphics[width=0.6\columnwidth]{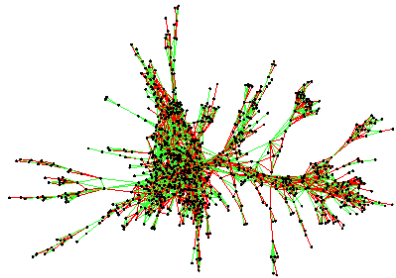}
\caption{(Color online) A sample network given by model C for $p = 0.6$. Positive and negative links are denoted by green (light grey) and red (dark grey) links, respectively.}
\label{modelBpic}
\end{centering}
\end{figure}

\noindent A typical network constructed by Model C is shown in (Fig. \ref{modelBpic}). The only free parameter is the probability of removal of the duplicated links, which is chosen to be $p = 0.6$. The epistatic signs are thoroughly mixed in this network, due to the equally probable assignment of positive and negative interactions between the target node and the duplicated node, and the ensuing sign convention of the duplicated links. This model permits the generation of two types of triangular motifs, namely pnn and ppp triangles (Fig. \ref{Bmotifs}) where p and n indicate respectively positive (buffering) and negative (aggravating) interactions. The underlying protein complex assumption is supported by the findings that pnn triangle motifs are encountered mostly between protein complexes separated by a negative edge suggesting a supportive coordination of function between them \cite{bandyopadhyay, casey}.

\begin{figure}[h!]
\centering
\mbox{\hspace{-5mm}
\subfigure{\includegraphics[width=0.27\textwidth]{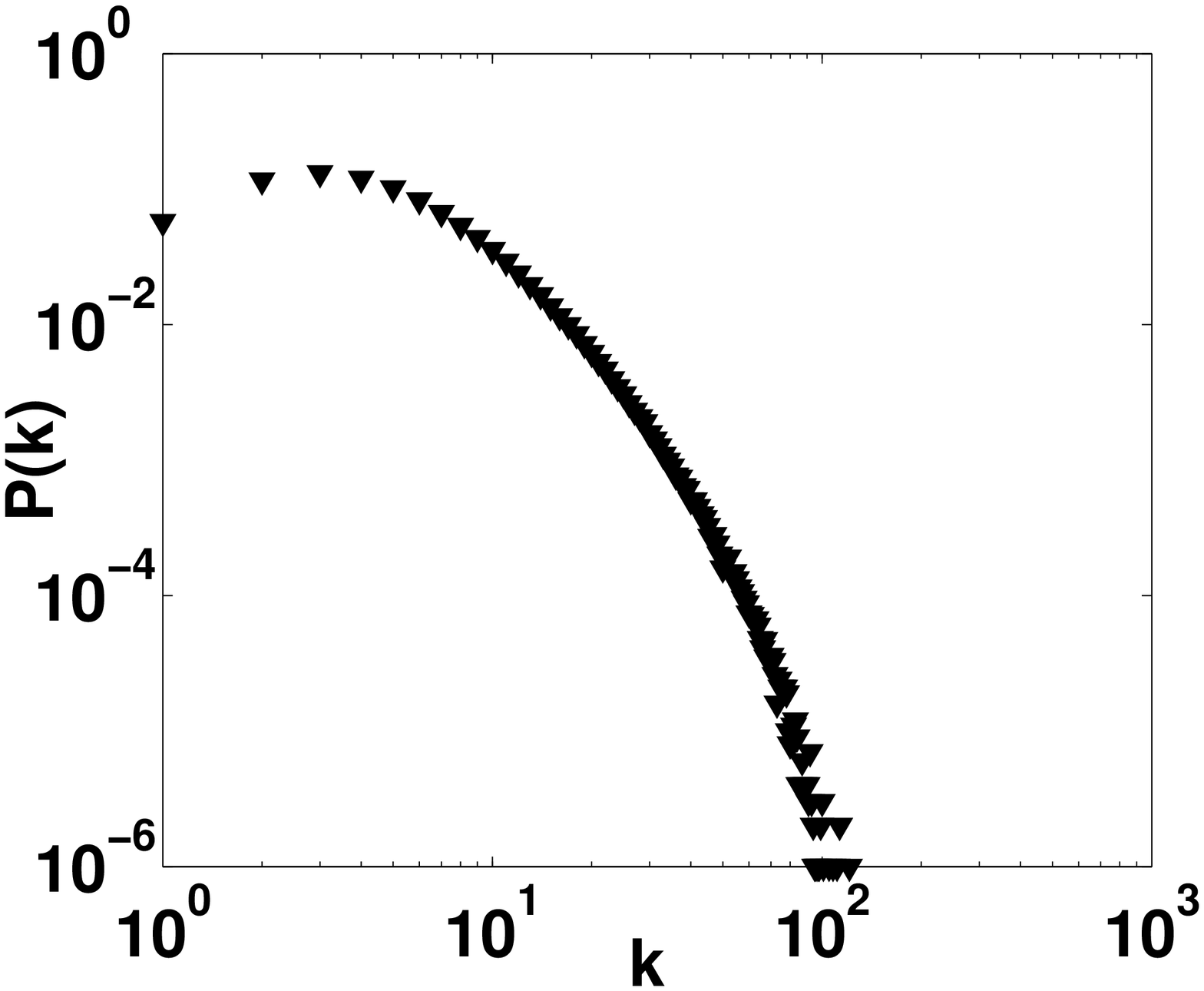}}
\hspace{-5mm}
\subfigure{\includegraphics[width=0.27\textwidth]{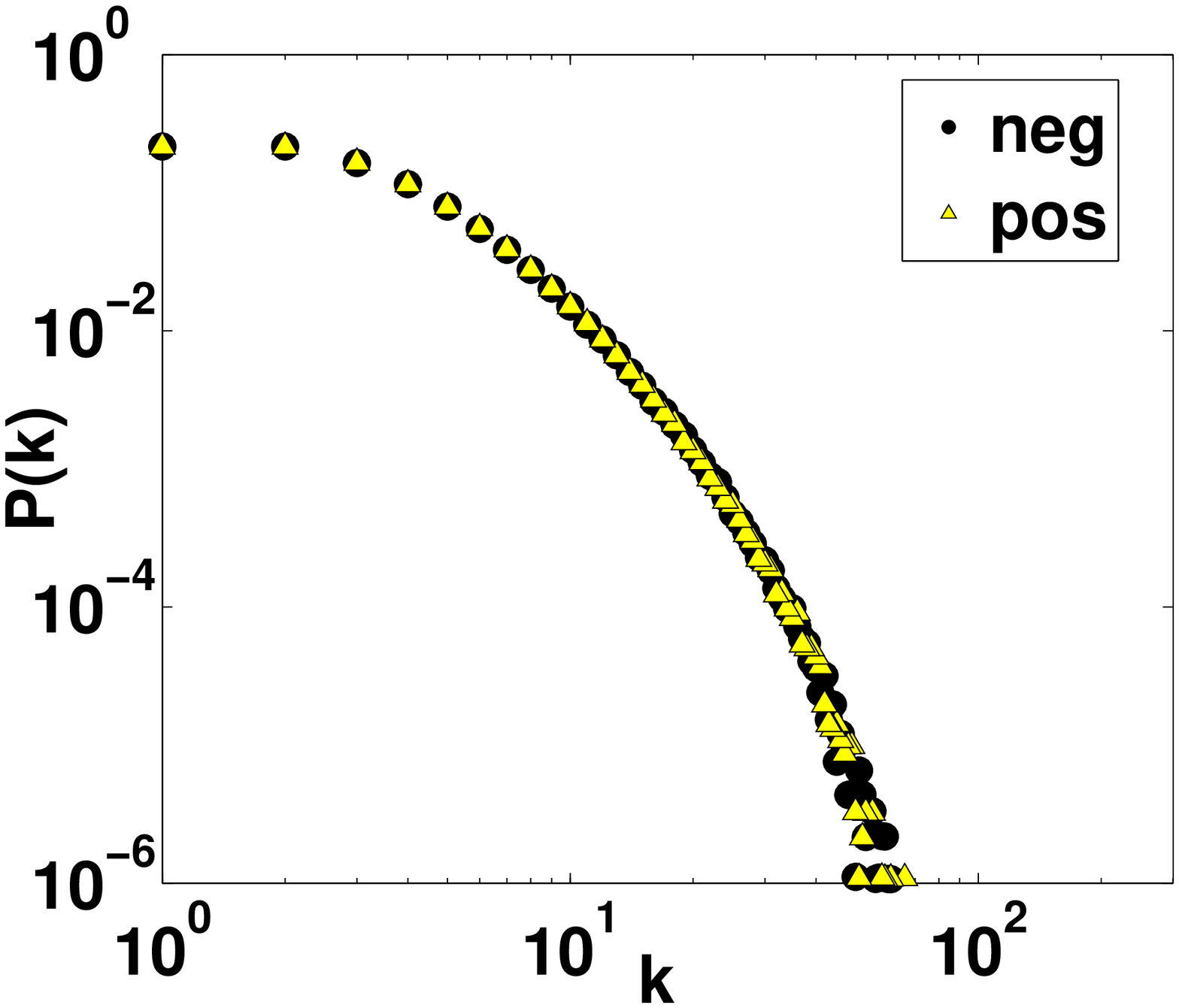}}}
\caption{(Color online) The total (left) and sign (right) degree distributions of model C for $p = 0.6$. Circles and triangles denote negative and positive links, respectively.}
\label{modelBdist}
\end{figure}

The degree distributions of this model are fat tailed as well despite the presence of some skewness in the power law distribution. The degree  distributions of interactions with different  signs follow the same trend (Fig. \ref{modelBdist}).

\begin{figure}[h!]
\begin{centering}
\vspace{-3mm}
\includegraphics[width=0.6\columnwidth]{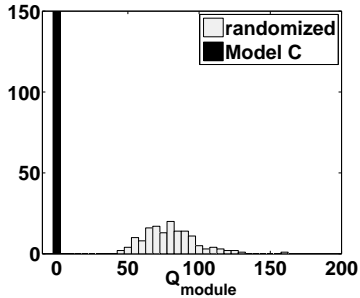}
\caption{The total monochromaticity violation number $Q_{module}$ distributions of model C networks (black) and their randomized counterparts (white) for 150 realizations of each. The parameter is chosen as $p = 0.6$}
\label{Bconflict}
\end{centering}
\end{figure}

An ensemble of 150 networks was used for the total conflict distribution. Compared to the other two models, this model is much more defined in terms of monochromaticity and gives a fully monochromatic solution (Fig. \ref{Bconflict}). The separation of the duplication divergence model from its randomized counterpart ($80.16 \pm 18.80$) is apparent. 

\section{Conclusion}
In conlusion in this paper we have investigated the modular structure and monochromatic clusterability of epistatic networks generated using stochastic models inspired by the duplication-divergence method in order to test for a neutral evolutionary hypothesis. We  have shown that these genetic interaction networks have total monochromatic violation score distributions that are close to zero and are statistically significantly different from their randomized counterparts. This fact implies that these networks comply with the monochromaticity feature observed in real genetic interaction data of model organisms such as the budding yeast.

Overall, the investigation of models based on the duplication-divergence mechanism is a good first step toward understanding the evolutionary origins of monochromaticity although it is important to stress that these mathematical models are very simplistic and further investigation into real biological networks is called for to extract information about what features are more relevant to monochromaticity in order to be able to tweak the details of these models. The parameter space of the three proposed models were explored before the $Q_{module}$ distributions were plotted and the free parameters were chosen so as to reproduce the general topological properties (scale-free degree distribution, average connectivity) of model organisms. Therefore it is fair to say that the results presented in this study constitute a â€œsnapshotâ€ of the models since modularity comes about in a relatively narrow range of parameters. Duplication-divergence does not account for the preference of these specific parameters. It is likely that they might be the result of some selection for sparse graphs in nature  \cite{gpwagner}.  So biological modularity cannot be explained solely by natural selection or neutral evolution but rather as the result of contributions from both -- an example being neutral mechanisms such as duplication-divergence whose parameters are tweaked by natural selection. The same is most likely valid for monochromaticity. In fact it is difficult to say that neutral evolution is the only explanation underlying it, although our findings with duplication-divergence models do corroborate its role.

One possible future direction to take, as far as assessing what the relevant structural characteristics in real biological networks are, would be to do a motif significance investigation on real data. Motifs are highly represented subgraphs in networks that occur statistically more frequently than they should in the randomized counterparts of networks  \cite{alon}. Although the immediate functional significance of individual motifs has been a disputed topic, it is still informative about the network structure.  Interesting parallels might be drawn between the motifs of real networks and implemented mathematical models. Model C in particular calls for such an analysis for different triangle types (4 overall). A comparison between these particular motifs in the yeast interaction network and networks generated by Model C can help improve the already promising ability of this model to reproduce monochromaticity in a biologically motivated way. An important point to note is that Model  C as it is presented in this work can produce two out of the four possible triangle motifs due to transitivity of signs. The two remaining motifs that are not produced by this model, namely nnn and ppn triangles, are mainly associated with synthetic lethal interactions and biochemical pathways, respectively \cite{casey}. Exploring the monochromaticity of models that result in these motifs might provide a more complete picture of the monochromaticity phenomenon in a wider scope of biological interactions.


\begin{thebibliography}{99}
\bibitem{Barabasi}
A.-L.  Barab\'asi and Z. Oltvai, 
Network biology: understanding the cell functional organization, 
{\em Nature Reviews} {\bf 5}, 101-113 (2004). 

\bibitem{Toroczkai}
E. Ben-Naim,H.  Frauenfelder  and A. Toroczkai 
{\em Complex Networks} Lecture Notes in Physics 650, (Springer-Verlag,2004).

\bibitem{Sneppen_book}
K. Sneppen  and G. Zocchi , 
{\em Physics in molecular biology} (Cambridge University Press, Cambridge,2005).

\bibitem{Bornholdt}
S. Bornholdt,
Less is More in Modelling Large Genetics Networks,
{\em Science} {\bf 310}, 449-451 (2005).

\bibitem{Reka}
R. Albert, Scale-free networks in cell biology, 
{\em Jour. of Cell Science} {\bf 118}, 4947-4957 (2005).

\bibitem{Kepes}
N. Guelzim, S. Bottani, P. Bourgine and F. K\'ep\'es,
Topological and causal structure of the yeast transcriptional regulatory network, 
{\em Nature Genetics } {\bf 31}, 60-63 (2002).

\bibitem{Alon}
U. Alon,
{\em An introduction to system biology: design principles of biological circuits}
(Chapman \& Hall, 2007).

\bibitem{Modular}
E. Ravasz, A. L. Somera, D. A.  Mongru, Z. N. Oltvai and A.-L. Barab\'asi, Hierarchical Organization of Modularity in Metabolic Networks, 
{\em Science} {\bf 297}, 1551-1555 (2002).

\bibitem{Santo}
S. Fortunato, Community detection in graphs, 
{\em Physics Reports} {\bf 486} 75-174 (2010).

\bibitem{PNAS}
G. Bianconi, P. Pin and M. Marsili,
{Assessing the relevance of node features for network structure},
{\em Proceedings of the National Academy of Science} {\bf 106}. 11433-11438 (2009).

\bibitem{Sole_modular}
R. V. Sol\'e and S. Valverde, J. R. Soc. Interface {\bf  5 }, 129 (2008).

\bibitem{segre}
D. Segr\`{e}, A. Deluna, G. M. Church, R. Kishony, {\it Nature Genetics} {\bf 37,} 77 (2005).
\bibitem{costanzo}
M. Costanzo {\it et al., Science} {\bf 327,} 425 (2010).

\bibitem{phillips}
P. C. Phillips,  {\it Nature Reviews Genetics} {\bf 9,} 855-867 (2008).

\bibitem{Uetz}
P. Uetz, L. Giot, G. Cagney, T. A. Mansfield, R. S. Judson, J. R. Knight, D. Lockshon, V. Narayan, M. Srinivasan, P. Pochart, A. Qureshi-Emili, Y. Li, B. Godwin, D. Conover, T. Kalbfleisch, G. Vijayadamodar, M. Yang, M. Johnston, S. Fields and J. M. Rothberg3,A comprehensive analysis of proteinâ€“protein interactions in Saccharomyces cerevisiae, 
{\em  Nature} {\bf 403}, 623-627 (2000).

\bibitem{Sneppen}
S. Maslov  and K. Sneppen   Specificity and stability in topology of protein networks, 
{\em Science} {\bf 296}, 910-913 (2002).

\bibitem{Laszlo_met}
H. Jeong, B. Tombor, R. Albert, Z.N. Oltvai and A.-L. Barab\'asi,
The large-scale organization of metabolic networks, 
{\em Nature} {\bf 407}, 651-654 (2000).











\bibitem{Marco}
M. Cosentino Lagomarsino, P. Jona, B. Bassetti and H. Isambert, PNAS {\bf 104} 5516 (2007).

\bibitem{hartwell}
L.H. Hartwell, J.J. Hopfield, S. Leibler, A.W. Murray, {\it Nature} {\bf 402}:C47-C52 (1999).
\bibitem{ravasz}
E. Ravasz, A. L. Somera, D. A. Mongru, Z. N. Oltvai and A.-L. Barabasi, {\it Science} {\bf 297,} 1551 (2002); 
\bibitem{ziv}
E. Ziv, M. Middendorf, C.H. Wiggins, {\it Phys. Rev. E} {\bf 71,} 046117 (2005).

\bibitem{baryshnikova}
A. Baryshnikova {\it et al., Nature Methods} {\bf 7,} 1017 (2010).
\bibitem{hsu}
C-H Hsu, T-Y Wang, H-T Chu, C-Y Kao, K-C Chen, {\it BMC Informatics} {\bf 12}(Suppl 13):S16 (2011).



\bibitem{sole}
R.V. Sol\'{e}, R. Pastor-Satorras, E. Smith, T.B. Kepler, {\it Adv. Complex Systems} {\bf 5,} 43 (2002).
\bibitem{vazquez}
A. V\'{a}zquez, A. Flammini, A. Maritan, A. Vespignani {\it Complexus} {\bf 1,} 38 (2003).
\bibitem{kim}
J. Kim, P. L. Krapivsky, B. Kahng, S. Redner, {\it Phys. Rev. E} {\bf 66,} 055101(R) (2002).

\bibitem{wagner}
A. Wagner, {\it Proc. Roy. Soc. London B} {\bf 270,} 457 (2003).


\bibitem{ohno} 
S. Ohno. {\it Evolution by gene duplication} Springer-Verlag, New York. (1970).


\bibitem{krapivsky}
I. Ispolatov, P. L. Krapivsky, A. Yuryev, {\it Phys. Rev. E} {\bf 71,} 061911 (2005).
\bibitem{sole2}
R. V. Sol\'{e}. P. Fernandez, {\it arXiv:q-bio.GN/0312032v1} (2003).

\bibitem{bandyopadhyay}
S. Bandyopadhyay, R. Kelley, N.J. Krogan, T. Ideker, {\it PLOS Computational Biology} {\bf 4,} 1000065 (2008).

\bibitem{casey}
F. Casey, N. Krogan, D.C. Shields, G. Cagney {\it BMC Systems Biology} {\bf 5,} 133 (2012).


\bibitem{gpwagner}
G. P. Wagner, M. Pavlicev, J. M. Cheverud, {\it Nature Reviews Genetics} {\bf 8,} 12 921-932 (2007).
\bibitem{alon}
R. Milo, S. Shen-Orr, S. Itzkovitz, N. Kashtan, D. Chklovskii, U. Alon, {\it Science} {\bf 298}:824-827 (2002). 













\end{thebibliography}
 \end{document}